\documentclass[%
superscriptaddress,
showpacs,
nofootinbib,
amsmath,amssymb,
prc,
twocolumn,
floatfix ]%
{revtex4-1}

\usepackage{color}
\usepackage{CJK}

\usepackage{graphicx}% Include figure files
\usepackage{dcolumn}% Align table columns on decimal point
\usepackage{bm}% bold math
\usepackage{hyperref}% add hypertext capabilities
\allowdisplaybreaks

\begin{document}

\begin{CJK*}{UTF8}{}
\title{Microscopic self-consistent description of induced fission: dynamical pairing degree of freedom}
\CJKfamily{gbsn}
\author{Jie Zhao (赵杰)}%
%\email{zhaojie@alumni.itp.ac.cn}
\affiliation{Center for Circuits and Systems, Peng Cheng Laboratory, Shenzhen 518055, China}
%\affiliation{Center for Quantum Computing, Peng Cheng Laboratory, Shenzhen 518055, China}
%\affiliation{Microsystem and Terahertz Research Center and Insititute of Electronic Engineering, 
%	China Academy of Engineering Physics, Chengdu 610200, Sichuan, China}
% \affiliation{Physics Department, Faculty of Science, University of Zagreb, Bijeni\v{c}ka Cesta 32,
%             Zagreb 10000, Croatia}
% \affiliation{CAS Key Laboratory of Theoretical Physics,
%              Institute of Theoretical Physics, Chinese Academy of Sciences, Beijing 100190, China}
\author{Tamara Nik\v{s}i\'c}%
%\email{tniksic@phy.hr}
\affiliation{Physics Department, Faculty of Science, University of Zagreb, Bijeni\v{c}ka Cesta 32,
        	      Zagreb 10000, Croatia}             
\author{Dario Vretenar}%
%\email{vretenar@phy.hr}
\affiliation{Physics Department, Faculty of Science, University of Zagreb, Bijeni\v{c}ka Cesta 32,
              Zagreb 10000, Croatia}
 \affiliation{ State Key Laboratory of Nuclear Physics and Technology, School of Physics, Peking University, Beijing 100871, China}            
%\CJKfamily{gbsn}              
%\author{Shan-Gui Zhou (周善贵)}%
%\email{sgzhou@itp.ac.cn}
 %\affiliation{CAS Key Laboratory of Theoretical Physics, Institute of Theoretical Physics, Chinese Academy of Sciences, Beijing 100190, China}
 %\affiliation{School of Physical Sciences, University of Chinese Academy of Sciences, Beijing 100049, China}
 %\affiliation{Center of Theoretical Nuclear Physics, National Laboratory of Heavy Ion Accelerator, Lanzhou 730000, China}
 %\affiliation{Synergetic Innovation Center for Quantum Effects and Application, Hunan Normal University, Changsha 410081, China}

\date{\today}

\begin{abstract}
The role of dynamical pairing in induced fission dynamics is investigated using the time-dependent generator coordinate method in the Gaussian overlap approximation, based on the microscopic framework of nuclear energy density functionals. A calculation of fragment charge yields for induced fission of $^{228}$Th is performed in a three-dimensional space of collective coordinates that, in addition to the axial quadrupole and octupole intrinsic deformations of the nuclear density, also includes an isoscalar pairing degree of freedom. It is shown that the inclusion of dynamical pairing has a pronounced effect on the collective inertia, the collective flux through the scission hyper-surface, and the resulting fission yields, reducing the asymmetric peaks and enhancing the contribution of symmetric fission, in better agreement with the empirical trend. 
\end{abstract}

\maketitle

\end{CJK*}

\bigskip

\section{Introduction~\label{sec:Introduction}}

Nuclear density functional theory (DFT) is the only microscopic framework that can be used over the entire table of nuclides in a self-consistent description of phenomena ranging from ground-state properties and collective excitations, to large-amplitude nucleonic motion, fission and low-energy collisions. In the case of nuclear fission \cite{Bender2020_JPG47-113002}, in particular, a fully quantum mechanical many-body model can be constructed starting from the time-dependent generator coordinate method (TDGCM)~\cite{Berger1991_CPC63-365}. In this approach the nuclear wave function is represented by a linear superposition of many-body generator states that are functions of collective coordinates. 
In most cases these coordinates parameterize the shape of the nuclear density. The Hill-Wheeler equation of motion determines the time evolution of the wave function in the restricted space of generator states \cite{Verriere2020_FP8-233}. 
By employing the Gaussian overlap approximation (GOA), the GCM Hill-Wheeler equation reduces to
a local, time-dependent Schr\"odinger equation in the space of collective coordinates. The microscopic input for the collective Schr\"odinger equation, that is, the nuclear potential and collective inertia, are determined by self-consistent mean-field calculations for a choice of the energy density functional (EDF) or effective interaction. The TDGCM+GOA method can be applied to the dynamics of induced fission, starting from the ground state and following the time evolution of collective degrees of freedom all the way to scission and the emergence of fission fragments. This framework has been very successfully implemented in a number of fission studies based on nonrelativistic Skyrme and Gogny 
functionals~\cite{Berger1991_CPC63-365,Verriere2020_FP8-233,Goutte2005_PRC71-024316,Regnier2016_CPC200-350,Regnier2018_CPC225-180,Regnier2016_PRC93-054611,Regnier2017_EPJWC146-04043,Verriere2017_EPJWC146-04034,Regnier2019_PRC99-024611,Zdeb2017_PRC95-054608}. These studies have investigated the dependence of the predicted fission dynamics on the choice of the EDF, initial conditions, form of the collective inertia, and the definition of scission configurations. 

Relativistic energy density functionals~\cite{Vretenar2005_PR409-101,Meng2006_PPNP57-470,Meng2016_WorldSci} have also been employed in the description of both spontaneous~\cite{Zhao2015_PRC92-064315,Zhao2016_PRC93-044315} and induced nuclear 
fission~\cite{Tao2017_PRC96-024319,Zhao2019_PRC99-014618,Zhao2019_PRC99-054613,Zhao2020_PRC101-064605}. The microscopic input for these studies is generated using either the multidimensionally constrained relativistic mean-field (MDC-RMF)~\cite{Lu2014_PRC89-014323} or the relativistic Hartree-Bogoliubov model 
(MDC-RHB)~\cite{Zhao2017_PRC95-014320}. By employing the TDGCM+GOA collective model, several interesting topics have been explored in this framework, such as the influence of static pairing correlations on fission yields, different approximations for the collective inertia tensor, and finite temperature effects. 

Most applications of the TDGCM to fission dynamics have considered a two-dimensional space of collective coordinates such as, for instance, quadrupole and octupole shape degrees of freedom. The recently developed computer code FELIX \cite{Regnier2016_CPC200-350,Regnier2018_CPC225-180} offers the possibility of solving the TDGCM+GOA equation for an arbitrary number of collective variables. In Ref.~\cite{Regnier2017_EPJWC146-04043} a preliminary calculation of induced fission dynamics of $^{240}$Pu isotope in the three-dimensional space of shape variables (quadrupole, octupole and hexadecupole intrinsic deformations) has been reported. In particular, this model can also be used for a quantitative analysis of the critical role of dynamical pairing correlations in the process of induced fission.

The importance of pairing correlations for both spontaneous and induced nuclear fission has been emphasized in a number of 
studies~\cite{Negele1978_PRC17-1098,Lojewski1999_NPA657-134,Sadhukhan2014_PRC90-061304,
Staszczak1985_PLB161-227,Zhao2016_PRC93-044315,Bernard2019_PRC99-064301,Lazarev1987_PS35-255,
Sadhukhan2016_PRC93-011304,Bulgac2016_PRL116-122504,Qiang2021_PRC103-L031304}.
For spontaneous fission it has been shown that the coupling between shape and pairing degrees of freedom has a pronounced effect on the calculated  fission lifetimes~\cite{Sadhukhan2014_PRC90-061304,Zhao2016_PRC93-044315}. 
In particular, when the gap parameter is considered as a dynamical variable, pairing correlations are generally enhanced thus reducing the effective inertia and the action integral along the fission path. This effect can significantly reduce the estimated spontaneous fission
lifetimes, and it has also been noted that pairing fluctuations can restore axial symmetry in the fissioning system.  A study of induced fission of $^{240}$Pu, using the microscopic time-dependent superfluid local density approximation (TD SLDA) \cite{Bulgac2016_PRL116-122504}, has shown that both shape and pairing modes determine the dynamics of the final stage of the fission process, from configurations close to the outer fission barrier to full scission.

The influence of ground-state (static) pairing correlations on charge yields and total kinetic energy of fission fragments for the case of induced fission of $^{226}$Th isotope was analyzed in Ref.~\cite{Tao2017_PRC96-024319} using the TDGCM+GOA framework. It has been shown that an increase of the strength of the pairing interaction, beyond the range determined by empirical pairing gaps obtained from the experimental masses of neighboring nuclei, 
reduces the asymmetric peaks and enhances the symmetric peak in charge yields distribution.
This is a very interesting result, and thus it is important to explore dynamical pairing correlations in induced fission. In this work we explicitly include the isoscalar pairing degree of freedom in the space of TDGCM+GOA collective coordinates, and perform the first realistic three-dimensional calculation of induced fission of $^{228}$Th. 
The theoretical framework and methods are reviewed in Sec.~\ref{sec:model}. The details of the calculation and principal results are discussed in
Sec.~\ref{sec:results}. Section~\ref{sec:summary} contains a short summary and outlook for future studies.

%%%%%%%%%%%%%%%%%%%%%%%%%%%%%%%%%%%%%%%%%%%%
\section{\label{sec:model}The TDGCM+GOA method}
%%%%%%%%%%%%%%%%%%%%%%%%%%%%%%%%%%%%%%%%%%%%
In the TDGCM+GOA framework induced fission is described as a slow adiabatic process determined by a small number of collective degrees of freedom. The initial step in modeling the fission of a heavy nucleus is a self-consistent mean-field (SCMF) calculation of the corresponding deformation energy surface as a function of few selected collective coordinates. Such a calculation provides the microscopic input, that is, the single-quasiparticle states, energies, and occupation factors, that determine the parameters of a local equation of motion for the collective wave function. 

As in our previous studies, here we use the point-coupling relativistic energy density functional
DD-PC1~\cite{Niksic2008_PRC78-034318} in the particle-hole channel, while pairing correlations are taken into account in the Bardeen-Cooper-Schrieffer (BCS) approximation by a separable
pairing force of finite range \cite{Tian2009_PLB676-44}:
\begin{equation}
V(\mathbf{r}_1,\mathbf{r}_2,\mathbf{r}_1^\prime,\mathbf{r}_2^\prime) = G_0 ~\delta(\mathbf{R}-
\mathbf{R}^\prime) P (\mathbf{r}) P(\mathbf{r}^\prime) \frac{1}{2} \left(1-P^\sigma\right),
\label{pairing}
\end{equation}
where $\mathbf{R} = (\mathbf{r}_1+\mathbf{r}_2)/2$ and $\mathbf{r}=\mathbf{r}_1- \mathbf{r}_2$
denote the center-of-mass and the relative coordinates, respectively. $P(\mathbf{r})$ reads 
\begin{equation}
P(\mathbf{r})=\frac{1}{\left(4\pi a^2\right)^{3/2}} e^{-\mathbf{r}^2/4a^2}.
\end{equation}
The parameters of the interaction were originally 
adjusted to reproduce the density dependence of the pairing gap in nuclear matter at the
Fermi surface computed with the D1S parameterization of the Gogny force~\cite{Berger1991_CPC63-365}.
To reproduce the empirical pairing gaps in the mass region considered in the present study, the strength parameters 
of the pairing force have been increased 
with respect to the original values by the following factors: $G_{n}/G_{0}=1.12$ and $G_{p}/G_{0}=1.08$ for 
neutrons and protons, respectively.

In the present analysis the self-consistent deformation energy surfaces are calculated using the multidimensionally constrained relativistic mean-field (MDC-RMF) model \cite{Lu2012_PRC85-011301R,Lu2014_PRC89-014323,Zhou2016_PS91-063008,Zhao2016_PRC93-044315}  with constraints on mass multipole moments  
$Q_{\lambda\mu} = r^{\lambda} Y_{\lambda\mu}$, and the particle-number dispersion operator $\Delta \hat{N}^{2} = \hat{N}^{2} - \langle \hat{N} \rangle^{2}$.
The Routhian is therefore defined as 
\begin{equation}
\label{eq:Routhian}
E^{\prime} = E_{\rm{RMF}} + \sum_{\lambda\mu} \frac{1}{2} C_{\lambda\mu} Q_{\lambda\mu} + \lambda_{2} \Delta \hat{N}^{2},
\end{equation}
where $E_{\rm RMF}$ denotes the total RMF energy that includes static BCS pairing correlations. 
The amount of dynamic pairing correlations can be controlled by the Lagrange multipliers 
$\lambda_{2\tau}$ ($\tau=n,p$),~\cite{Vaquero2011_PLB704-520,Vaquero2013_PRC88-064311,Sadhukhan2014_PRC90-061304,Zhao2016_PRC93-044315}.
To reduce the number of collective degrees of freedom and, therefore, the considerable computational task, here we only consider isoscalar 
dynamical pairing; $\lambda_{2n} = \lambda_{2p} \equiv \lambda_{2}$ is employed as the collective coordinate.
The nuclear shape is parameterized by the deformation parameters
\begin{equation}
 \beta_{\lambda\mu} = {4\pi \over 3AR^\lambda} \langle Q_{\lambda\mu} \rangle,
\end{equation}
with $R=1.2~A^{1/3}$ fm.
The shape is assumed to be invariant under the exchange of the $x$ and $y$ axes, 
and all deformations $\beta_{\lambda\mu}$ with even $\mu$ can be included simultaneously.
The constrained RMF+BCS equations are solved by an expansion in the 
axially deformed harmonic oscillator (ADHO) basis~\cite{Gambhir1990_APNY198-132}.
In the present study calculations have been performed 
in an ADHO basis truncated to $N_f = 20$ oscillator shells.

The dynamics of the fission process is thus governed by a local, 
time-dependent Schr\"odinger-like equation 
in the space of collective coordinates $\bm{q}$:
\begin{equation}  
i\hbar \frac{\partial}{\partial t}g(\bm{q},t) = \hat{H}_{\rm coll} (\bm{q}) g(\bm{q},t) , 
\label{eq:TDGCM}
\end{equation}
where $g(\bm{q},t)$ is the complex wave function of the collective variables $\bm{q}$ and time $t$. 
The collective Hamiltonian $\hat{H}_{\rm coll} (\bm{q})$ 
\begin{align}
\hat{H}_{\rm coll} (\bm{q}) = - {\hbar^2 \over 2} 
	\sum_{ij} {\partial \over \partial q_{i}}  B_{ij}(\bm{q}) {\partial \over \partial q_{j}} + V(\bm{q}),
\label{eq:Hcoll2}
\end{align}
governs the time-evolution of the nuclear wave function from an initial state at equilibrium deformation, up to scission and the formation of fission fragments.
$B_{ij}(\bm{q})$ and $V(\bm{q})$ denote the inertia tensor and collective potential, respectively, 
that are computed using the self-consistent solutions (total energy, single-quasiparticle states and occupation factors) for the RMF+BCS deformation energy surface.
Here we assume axial symmetry with respect to the axis along which the two fragments eventually separate, 
and consider the three-dimensional (3D) collective space of quadrupole $\beta_{2}$ and octupole $\beta_{3}$ deformation parameters, 
and the dynamical pairing coordinate $\lambda_{2}$.
The inertia tensor is the inverse of the mass tensor, that is, $B_{ij}(\bm{q}) =(\mathcal{M}^{-1})_{ij}(\bm{q})$.
The mass tensor is calculated using the adiabatic time-dependent Hartree-Fock-Bogoliubov (ATDHFB) method in the cranking approximation~\cite{Baran2011_PRC84-054321}: 
%
%The adiabatic time-dependent Hartree-Fock-Bogoliubov (ATDHFB) method is applied in both the non-perturbative and perturbative cranking approximations to the calculation of the 
%mass tensor.
%In the cranking approximation the mass tensor takes the form \cite{Baran2011_PRC84-054321} 
\begin{equation}
\label{eq:npmass}
\mathcal{M}_{ij}^{C} = {\hbar^2 \over 2 \dot{q}_i \dot{q}_j}
    \sum_{\mu\nu} {F^{i*}_{\mu\nu}F^{j}_{\mu\nu} + F^{i}_{\mu\nu}F^{j*}_{\mu\nu}
    \over E_{\mu} + E_{\nu}},
\end{equation}
where
\begin{equation}
\label{eq:fmatrix}
{F^{i} \over \dot{q}_{i}}  
  =  U^\dagger {\partial\rho \over \partial q_{i}} V^* 
    + U^\dagger {\partial\kappa \over \partial q_{i}} U^*
    - V^\dagger {\partial\rho^* \over \partial q_{i}} U^*
    - V^\dagger {\partial\kappa^* \over \partial q_{i}} V^*\;.
\end{equation}
$U$ and $V$ are the self-consistent Bogoliubov matrices, and $\rho$ and $\kappa$ are 
the corresponding particle and pairing density matrices, respectively.
The derivatives of the densities are calculated using the Lagrange three-point formula for 
unequally spaced points~\cite{Yuldashbaeva1999_PLB461-1,Baran2011_PRC84-054321}.
The cranking expression Eq.~(\ref{eq:npmass}) can be further simplified in the perturbative 
approach~\cite{Brack1972_RMP44-320,Nilsson1969_NPA131-1,Girod1979_NPA330-40,Bes1961_NP28-42,
Sobiczewski1969_NPA131-67}, and this leads to the perturbative cranking mass tensor: 
\begin{equation}
\label{eq:pmass}
\mathcal{M}^{Cp} = \hbar^2 {\it M}_{(1)}^{-1} {\it M}_{(3)} {\it M}_{(1)}^{-1}, 
\end{equation}
where 
\begin{equation}
\label{eq:mmatrix}
\left[ {\it M}_{(k)} \right]_{ij} = \sum_{\mu\nu} 
    {\left\langle 0 \left| \hat{Q}_i \right| \mu\nu \right\rangle
     \left\langle \mu\nu \left| \hat{Q}_j \right| 0 \right\rangle
     \over (E_\mu + E_\nu)^k}.
\end{equation}
$|\mu\nu\rangle$ are the two-quasiparticle states with the  corresponding quasiparticle energies $E_\mu$ and $E_\nu$.
The details of the derivation of cranking formulas for the mass tensor can be found in 
Ref.~\cite{Baran2011_PRC84-054321}. 

At this point we introduce an approximation that is not entirely consistent, but is necessary to 
reduce the computational task and stabilize the time-evolution of the collective state. 
Because we consider the particle-number dispersion operator $\Delta \hat{N}^{2} = \hat{N}^{2} - \langle \hat{N} \rangle^{2}$, the pairing part of the 
mass tensor must be calculated using the non-perturbative cranking expression Eq.~(\ref{eq:npmass}).  
In the recent study of the differences between the perturbative and non-perturbative ATDHFB collective masses in the TDGCM+GOA description of induced fission \cite{Zhao2020_PRC101-064605}, using the axial quadrupole and octupole intrinsic deformation as dynamical variables, we have shown that the structure of non-perturbative collective masses is much more complex due to changes in the intrinsic shell structure across the deformation energy surface, and is characterized by pronounced isolated peaks located at single-particle level crossings near the Fermi surface. It has been shown that the choice of non-perturbative cranking collective mass leads to a reduction of symmetric charge yields and, generally, to a better agreement with data. Even though in the analysis of Ref.~\cite{Zhao2020_PRC101-064605} both non-perturbative and perturbative mass tensors were used  in modeling induced fission dynamics, in the present 3D study the number of mesh points required to accurately calculate all the non-perturbative collective masses becomes
prohibitively large for the available computational resources. The reason is the occurrence of pronounced peaks in the collective masses related to single-particle level crossings near the Fermi surface, and the corresponding abrupt changes of the occupation factors of single-particle configurations. 
This would require a major refinement of the grid, leading to very large number of mesh points and possible instabilities. Therefore, to be able to quantitatively analyze the effect of dynamical pairing on induced fission, here we use the perturbative cranking expression Eq.~(\ref{eq:pmass}) to calculate the elements of the mass tensor that correspond to the quadrupole and octupole deformations while, as noted above, the pairing element of the mass tensor must be computed using the non-perturbative cranking formula. With the indices $1$, $2$ and $3$ corresponding to the $\beta_{20}$, $\beta_{30}$, and $\lambda_{2}$ collective coordinates, the following elements of the mass tensor are used: $\mathcal{M}^{Cp}_{11}$, $\mathcal{M}^{Cp}_{12}$, $\mathcal{M}^{Cp}_{22}$, $\mathcal{M}^{C}_{33}$,  
and we neglect the coupling terms $\mathcal{M}_{13}$ and $\mathcal{M}_{23}$. 

To model the fission dynamics we follow 
 the time-evolution of an initial wave packet $g(\bm{q},t=0)$ ($\bm{q} \equiv \{\beta_2,\beta_3,\lambda_2\}$), built 
as a Gaussian superposition of the quasi-bound states $g_k$, 
\begin{equation}
g(\bm{q},t=0) = \sum_{k} \exp\left( { (E_k - \bar{E} )^{2} \over 2\sigma^{2} } \right) g_{k}(\bm{q}),
\label{eq:initial-state}
\end{equation}
where the value of the parameter $\sigma$ is set to 0.5 MeV. The collective states $\{ g_{k}(\bm{q}) \}$ 
are solutions of the stationary eigenvalue equation in which the original collective potential $V(\bm{q})$ is replaced by a 
new potential $V^{\prime} (\bm{q})$ that is obtained by extrapolating the inner potential barrier with a quadratic form. 
The mean energy $\bar{E}$ in Eq.~(\ref{eq:initial-state}) is then adjusted iteratively in 
such a way that $\langle g(t=0)| \hat{H}_{\rm coll} | g(t=0) \rangle = E_{\rm coll}^{*}$, and this 
average energy $E_{\rm coll}^{*}$ is set to be 1 MeV above the fission barrier.
The TDGCM+GOA Hamiltonian of Eq.~(\ref{eq:Hcoll2}), with the original collective potential 
$V(\bm{q})$, propagates the initial wave packet in time.

The time propagation is modeled using the TDGCM+GOA computer 
code FELIX (version 2.0)~\cite{Regnier2018_CPC225-180}.
The time step is $\delta t=5\times 10^{-4}$ zs (1 zs $= 10^{-21}$ s), and the charge and mass 
distributions are calculated after $4\times10^{4}$ time steps, which correspond to 20 zs.
As in our recent calculations of Refs.~\cite{Tao2017_PRC96-024319,Zhao2019_PRC99-014618,Zhao2019_PRC99-054613,Zhao2020_PRC101-064605}, 
the parameters of the additional imaginary absorption potential that takes into account the escape 
of the collective wave packet  in the domain outside the region of calculation \cite{Regnier2018_CPC225-180} are: 
the absorption rate $r=20\times 10^{22}$ s$^{-1}$ and the width of the absorption band $w=1.0$.

The collective space is divided into an inner region with a single nuclear density distribution, 
and an external region that contains two separated fission fragments. 
The scission hyper-surface that divides the inner and external regions is determined by the 
expectation values of the 
Gaussian neck operator $\displaystyle \hat{Q}_{N}=\exp[-(z-z_{N})^{2} / a_{N}^{2}]$, 
where $a_{N}=1$ fm and $z_{N}$ is the position of the neck~\cite{Younes2009_PRC80-054313}.
We define the pre-scission domain by $\langle \hat{Q}_{N} \rangle>3$, and consider the frontier of this domain as the scission surface.
 The flux of the probability current through this
hyper-surface provides a measure of the probability of observing a given pair of fragments at time $t$.
Each infinitesimal surface element is associated with a given pair of fragments $(A_L, A_H)$, where $A_L$ and $A_H$ denote the 
lighter and heavier fragments, respectively.
The integrated flux $F(\xi,t)$ for a given surface element $\xi$ is defined as \cite{Regnier2018_CPC225-180}
\begin{equation}
F(\xi,t) = \int_{t_0}^{t} dt^\prime \int_{\bm{q} \in \xi} \bm{J}(\bm{q}, t^\prime) \cdot d\bm{S}, 
\label{eq:flux}
\end{equation}
where $\bm{J}(\bm{q}, t)$ is the current
\begin{align}
\label{eq:current}
J_{k}(\bm{q},t) = \hbar \sum_{l} B_{kl}(\bm{q}) {\mathrm{Im}}\left(g^* \frac{\partial g}{\partial q_{l}} \right).
\end{align}
The yield for the fission fragment with mass $A$ is defined by 
\begin{equation}
Y(A) \propto \sum_{\xi \in \mathcal{A}} \lim_{t \rightarrow \infty} F(\xi,t).
\end{equation}
The set $\mathcal{A}(\xi)$ contains all elements belonging to the scission hyper-surface such that one of the fragments has mass number $A$.
The charge yields are obtained by convoluting the raw flux with a Gaussian function of the number of particles  \cite{Regnier2016_PRC93-054611,Zhao2019_PRC99-014618}, with a width of 1.6 units.

%----------------------------------------------------------------------------------------------------------------------
\section{\label{sec:results}Results and discussion}
%----------------------------------------------------------------------------------------------------------------------
As an illustrative example, the fission of $^{228}$Th is considered. For this nucleus the charge distribution of fission fragments exhibits a coexistence of symmetric and asymmetric peaks~\cite{Schmidt2000_NPA665-221}.
In the first step a large scale MDC-RMF calculation is performed to generate the potential energy surface, 
single-nucleon wave functions and occupation factors in the $(\beta_2,\beta_3,\lambda_2)$ collective space.  
The intervals for the values of the collective variables are:  $-1 \le \beta_2 \le 7$ with   
a step $\Delta \beta_2 = 0.04$; $0 \le \beta_3 \le 3.5$ with a step $\Delta \beta_3 =0.05$;
and $-0.1 \le \lambda_2 \le 2.0$ with a step $\Delta \lambda_2 = 0.1$.
The relativistic energy density functional DD-PC1~\cite{Niksic2008_PRC78-034318} is used in the particle-hole channel, and  
particle-particle correlations are described by the separable pairing force (\ref{pairing}) in the BCS approximation.

\begin{figure}[!]
\centering
 \includegraphics[width=0.45\textwidth]{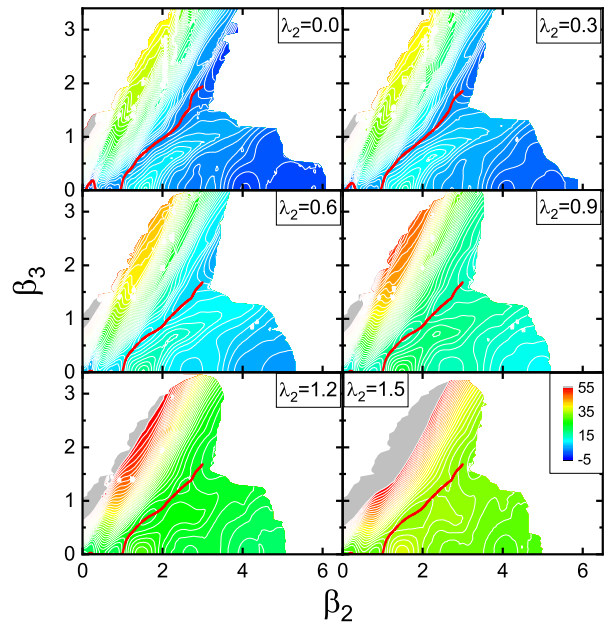}
 \caption{(Color online)~\label{fig:3DPES}%
 2D projections of the deformation-energy manifold of $^{228}$Th on the quadrupole-octupole axially symmetric plane, calculated with the RMF+BCS model based on the functional DD-PC1, 
 for selected values of the pairing coordinate $\lambda_{2}$. 
 Contours join points on the surface with the same energy, and the separation between 
 neighboring contours is 1 MeV.  The red curves denote static fission paths of minimum 
 energy for each value of $\lambda_2$.
}
\end{figure}

Figure \ref{fig:3DPES} displays the 2D projections of the 3D deformation energy surface of $^{228}$Th on the collective plane $(\beta_{2},\beta_{3})$, 
for several values of the collective coordinate $\lambda_2$. Note that the value $\lambda_2=0$  corresponds to static BCS 
pairing, while positive $\lambda_2$ equates to enhanced pairing correlations. Only configurations with $Q_N\ge 3$ are plotted, 
and the frontier of this domain determines the scission contour. The red curves correspond to static fission paths of minimum total energy. The deformation surfaces for $\lambda_2=0$ and $\lambda_2=0.3$
are very similar, with a pronounced ridge separating the asymmetric and symmetric fission valleys. This ridge decreases with increasing values of the pairing coordinate $\lambda_2$.  The scission contour for $\lambda_2=0$ starts from an elongated symmetric
point at $\beta_{2}\approx 6$, and evolves to a minimal elongation $\beta_{2}\approx 3$ as reflection asymmetry increases. For larger
values of $\lambda_2$ the scission contour is not modified significantly, although the starting point of the scission contour on the quadrupole axis shifts to
smaller elongations at $\beta_{2}\approx 5$.
%----
%-----

\begin{figure}[!]
\centering
 \includegraphics[width=0.45\textwidth]{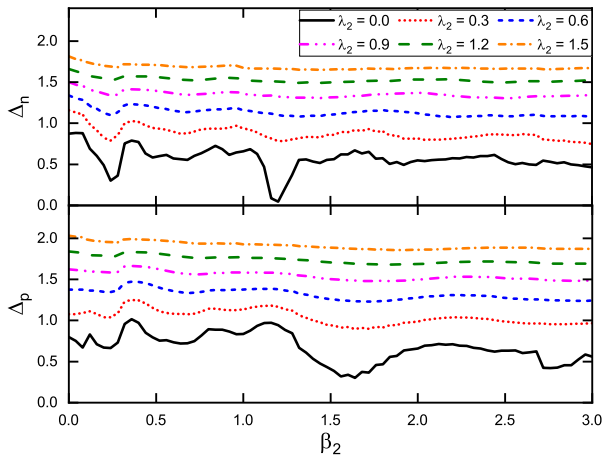}
\caption{(Color online)~\label{fig:Delta}%
Pairing gaps for neutrons $\Delta_{n}$ (upper panel) and protons $\Delta_{p}$ (lower panel)
along the static fission path, as functions of the axial quadrupole deformation, for selected values of the isoscalar pairing collective coordinate 
$\lambda_{2}$. 
}
\end{figure}

To illustrate the dynamical effect on the pairing correlations, in Fig.~\ref{fig:Delta}
we display the values of the neutron (upper panel) and proton (lower panel) pairing gap along the
static fission paths, as functions of the quadrupole coordinate $\beta_{2}$, for several values of the isoscalar pairing collective coordinate $\lambda_2$ 
(cf. Fig.~\ref{fig:3DPES}). In the interval of values of $\lambda_2$ considered here, the values of 
the pairing gaps increase by a factor $\approx 2-3$ and, characteristically, for $\lambda_2 \ge 0.6$ 
all traces of the underlying shell effects along the static fission path vanish.

%----
%-----

\begin{figure}[!]
\centering
 \includegraphics[width=0.45\textwidth]{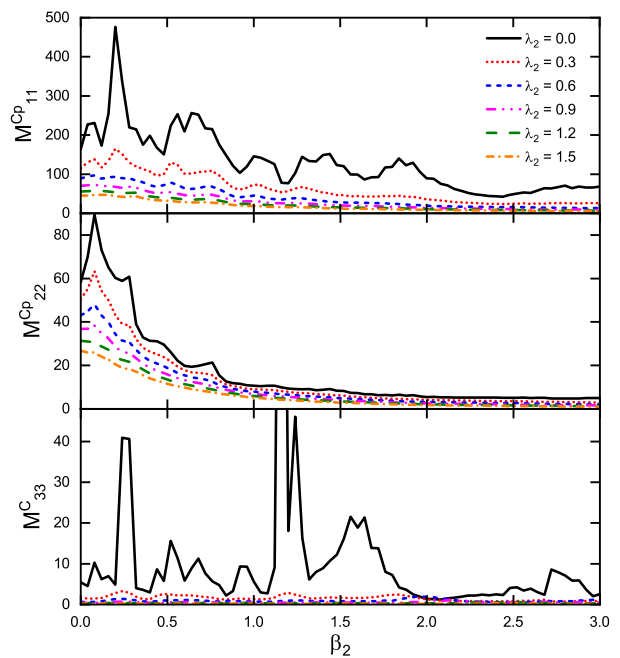}
\caption{(Color online)~\label{fig:Mass}%
Perturbative cranking masses $\mathcal{M}^{Cp}_{11}$, $\mathcal{M}^{Cp}_{22}$, 
and the non-perturbative cranking mass $\mathcal{M}^{C}_{33}$ (in $\hbar^{2}$ MeV$^{-1}$) 
along the static fission path for several values of $\lambda_{2}$. 
}
\end{figure}

In Fig.~\ref{fig:Mass} we plot the values of the perturbative cranking collective masses $M^{Cp}_{11}$, $M^{Cp}_{22}$, and the non-perturbative cranking mass $M^{C}_{33}$,
along the static fission path as functions of the quadrupole deformation, for different pairing collective coordinates $\lambda_2$. As noted above, the indices $1$, $2$ and $3$ refer to the $\beta_{2}$, $\beta_{3}$ and $\lambda_2$ coordinates, respectively. The collective masses $M^{Cp}_{11}$ and $M^{Cp}_{22}$ exhibit a more complex structure for
smaller values of deformations $\beta_{2}<1$ and decrease for larger deformations. One notices that $M^{Cp}_{22}$ exhibits a sharper decrease for $\beta_{2}>1$, compared to $M^{Cp}_{11}$.
The behaviour of the $M^{C}_{33}$ collective mass is much more complex, due to the fact that it has been calculated using the non-perturbative cranking formula Eq.~(\ref{eq:npmass}). This leads to the appearance of prominent peaks at the locations of single-particle level crossing near the Fermi surface. However, in general also $M^{C}_{33}$ 
decreases for large deformations and the peaks are less pronounced. 
As the value of the dynamical pairing coordinate $\lambda_2$ increases, all three collective masses 
decrease and the underlying shell effects gradually vanish. This effect is
consistent with the results shown in Fig. 6 of Ref.~\cite{Tao2017_PRC96-024319}.

%----
\begin{figure}[!]
\centering
 \includegraphics[width=0.45\textwidth]{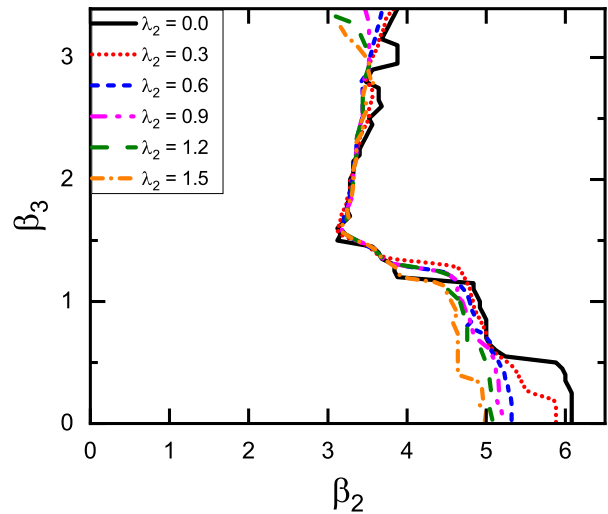}
\caption{(Color online)~\label{fig:SciLine}%
The scission controur of $^{228}$Th in the ($\beta_{2}, \beta_{3}$) deformation plane for several values of  the collective pairing coordinate$\lambda_{2}$ . 
}
\end{figure}

%-----
The scission contours in the ($\beta_{2}, \beta_{3}$) plane are shown in Fig.~\ref{fig:SciLine} for 
several collective pairing coordinates $\lambda_{2}$. The contours are generally not very different, especially for asymmetric fission. In particular, scission points that are close to the static fission path are not sensitive to dynamical pairing. For larger values of $\lambda_2$, however, the scission contour is shifted 
towards smaller quadrupole deformations $\beta_{2}$ values for nearly symmetric fission. 
%------------
%-----

\begin{figure}[!]
\centering
 \includegraphics[width=0.45\textwidth]{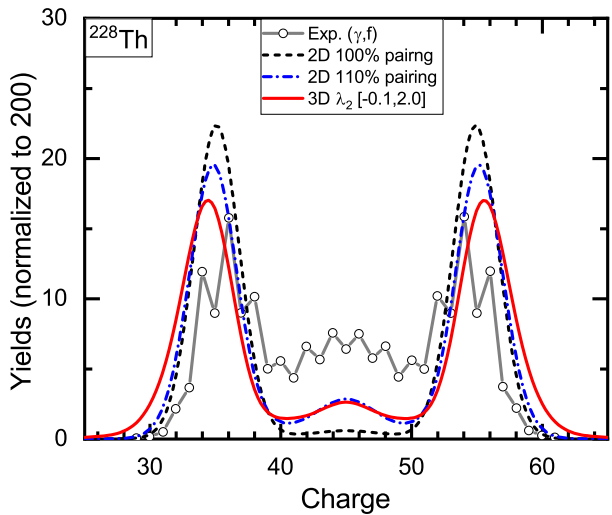}
 \caption{(Color online)~\label{fig:Zyields}%
 Charge yields for induced fission of $^{228}$Th, calculated in the 3D collective space 
 built from the deformation $\beta_{2}$, $\beta_{3}$ and dynamical pairing $\lambda_2$ coordinates (solid red curve). The yields are shown in comparison to the results obtained in the 2D space of 
shape degrees of freedom $\beta _{2}$ and $\beta_{3}$, with static pairing correlations adjusted to 
empirical ground-state pairing gaps (100\% pairing strength), and enhanced by ten percent 
(110\% pairing strength). The data for photo-induced fission correspond to photon energies in the interval 8-14 MeV,  and peak value of $E_{\gamma} = 11$ MeV~\cite{Schmidt2000_NPA665-221}.
}
\end{figure}

In Fig.~\ref{fig:Zyields} we compare the theoretical predictions for the charge yields with  the
data for photo-induced fission of $^{228}$Th. The 3D calculation employs the collective space built from the deformation $\beta_{2}$, $\beta_{3}$ and pairing $\lambda_2$ coordinates, while the 2D calculation includes only the shape degrees of freedom $\beta_{2}$ and $\beta_{3}$, and static pairing correlations. 
As in our study of Ref.~\cite{Tao2017_PRC96-024319}, the calculation in the 2D collective space 
corresponds to normal and enhanced static pairing, that is, to 100\% and 110\% of the normal pairing strength, determined by the empirical pairing gaps. The theoretical predictions follow the general trend of the data, except that our model obviously cannot reproduce the odd-even staggering of the experimental charge
yields. The calculation that includes only the 2D collective space with a static pairing strength 
adjusted to empirical ground-state pairing gaps in this mass region (100\%),  
predicts yields that are entirely dominated by asymmetric fission with peaks at $Z=35$ and $Z=55$. 
By increasing static pairing (110\%), the asymmetric peaks are reduced and a contribution of symmetric fission develops, but not strong enough to reproduce the data. It is interesting to notice that a very similar distribution of charge yields is predicted by the 3D model calculation that includes dynamical pairing. On a quantitative level, even the 3D calculation does not completely reproduce the experimental yields. 
The model predicts tails of the asymmetric peaks that are not seen in experiment,  
and thus fails to quantitatively match the symmetric contribution. It has to be noted, however, that in the present study the collective potential and inertia tensor have been calculated at zero temperature. In our recent study of finite temperature effects in TDGCM+GOA \cite{Zhao2019_PRC99-014618}, a calculation of induced fission of $^{226}$Th has shown that, although the model can qualitatively 
reproduce the empirical triple-humped structure of the fission charge and mass distributions already at 
zero temperature, the position of the asymmetric peaks 
and the symmetric-fission yield can be described much better when the potential and collective mass are determined at a temperature that approximately corresponds to the internal excitation energy of the fissioning nucleus. 

%----
%-----

\begin{figure}[!]
\centering
 \includegraphics[width=0.45\textwidth]{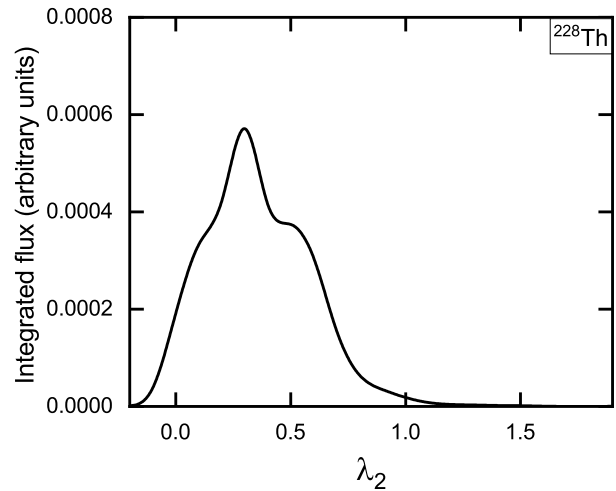}
\caption{\label{fig:flux}%
Time-integrated collective flux $B(\lambda_2)$ Eq.~(\ref{eq:Ylam2}) through the scission contour, as a function of the pairing collective coordinate $\lambda_2$.
}
\end{figure}

Finally, to illustrate the effect of dynamical pairing on the flux of the probability current through the scission hyper-surface, in Fig.~\ref{fig:flux} we plot the time-integrated flux through the scission contour 
in the $(\beta_{2},\beta_{3})$ plane, for a given value of the pairing collective coordinate $\lambda_2$
 \begin{equation}
 \label{eq:Ylam2}
B(\lambda_2) \propto \sum_{\xi \in \mathcal{B}}{\lim_{t\to\infty}F(\xi,\lambda_2,t)}.
\end{equation}
The set $\mathcal{B}(\xi \equiv \beta_2, \beta_3)$ contains all elements of the scission contour with a  given value $\lambda_2$. Even though it appears that dynamical pairing does not significantly modify the scission contour (cf. Fig.~\ref{fig:SciLine}), nevertheless its effect on the collective flux and, therefore, on the occurrence of fission, is remarkable. 
For negative values of $\lambda_2$, that is, for correlations weaker than static pairing at $\lambda_2 = 0$, the flux rapidly decreases to zero. For positive values of $\lambda_2$ the flux exhibits a steep increase and a prominent peak at $\lambda_2\approx 0.3$. Note that this value corresponds to an increase of $\approx 20\%$ with respect to the static proton and neutron pairing gaps (cf. Fig.~\ref{fig:Delta}). The collective flux through the scission contour weakens with a further increase of pairing, and eventually vanishes for $\lambda_2 > 1$. 
%------------

%-----

%
\section{\label{sec:summary}Summary}
The influence of dynamical pairing degrees of freedom on induced fission has been investigated in a unified theoretical framework based on the generator coordinate method (GCM) with the Gaussian overlap approximation (GOA). In an illustrative calculation of fragment charge yields for induced fission of $^{228}$Th, the collective potential and inertia tensor have been computed using the self-consistent multidimensionally constrained relativistic mean field model, based on the energy density functional DD-PC1, and with pairing correlations treated in the BCS approximation with a separable pairing force of finite range. The fission fragment charge distributions are obtained by propagating the initial collective state in time with the time-dependent GCM+GOA. The flux of the probability current through the scission hyper-surface determines the probability of observing a given pair of fragments. 

In this work, for the first time, the dynamics of induced fission has been consistently described in a three-dimensional space of collective coordinates that, in addition to the axial quadrupole and octupole intrinsic deformations of the nuclear density, also includes an isoscalar pairing degree of freedom. A number of studies has already demonstrated the importance of dynamical pairing for the calculation of spontaneous fission lifetimes, and static pairing correlations for modeling induced fission. As this work has also shown, a much more difficult problem is the inclusion of dynamical pairing degrees of freedom in a time-dependent description of induced fission. One expects, of course, that the model becomes more realistic as the Hilbert space of collective coordinates is expanded. 
However, as the present analysis has illustrated, it can be difficult to numerically stabilize the time-evolution of the fissioning system when shape and pairing collective coordinates are considered simultaneously in a three-dimensional calculation. Here it was necessary to compute the collective inertia using different cranking approximations for the shape and pairing degrees of freedom. Nevertheless, this approach enables a qualitative study of the effect of dynamical pairing on induced fission. 

This initial study has clearly demonstrated the important effect that dynamical pairing correlations have on the induced-fission fragment distribution. In particular, the charge distribution of fission fragments of $^{228}$Th is characterized by symmetric and asymmetric peaks, but this structure cannot be reproduced in a two-dimensional calculation that only includes shape collective variables. In that case, and with a static pairing strength adjusted to empirical ground-state pairing gaps, the calculated yields correspond to a completely asymmetric fission. Only by artificially increasing the static pairing correlations or, more naturally, by including the dynamical pairing degree of freedom in the three-dimensional calculation, the asymmetric peaks get reduced and a contribution of symmetric fission develops in agreement with the empirical trend. It is also interesting to note that the time-integrated collective flux through the scission contour in the $(\beta_{2},\beta_{3})$ plane, exhibits a characteristic functional dependence on the pairing collective coordinate, with a prominent peak at a value that correspond to an increase of $\approx 20\%$ with respect to the static pairing gaps. 

Future advances in computational capabilities will open the possibility of more quantitative applications of multi-dimensional TDGCM+GOA to fission dynamics. An immediate task will be to consider shape and pairing degrees of freedom on an equal footing, and consistently compute the corresponding collective inertia tensor and metric. A more challenging problem is to include dynamical pairing degrees of freedom in recently developed fission models that attempt to incorporate restoration of symmetries broken by the intrinsic densities in constrained mean-field calculations (rotational, reflection, and particle number symmetry) \cite{Marevic2020_PRL125-102504,Verriere2019_PRC100-024612,Verriere2021_PRC103-054602}.

%%%%%%%%%%%%%%%%%%%%%%%%%%%%%%%%%%%%%%%%%%
\bigskip
%---------------------------------------------------------
\acknowledgements
This work has been supported in part by the QuantiXLie Centre of Excellence, a project co-financed by the Croatian Government and European Union through the European Regional Development Fund - the Competitiveness and Cohesion Operational Programme (KK.01.1.1.01.0004) and the Croatian Science Foundation under the project Uncertainty quantification
within the nuclear energy density framework (IP-2018-01-5987).
It has also been supported by the National Natural Science Foundation of China under Grant No. 12005107 and No. 11790325.
%Calculations have been performed in part at the HPC Cluster of KLTP/ITP-CAS and the Supercomputing Center,
%Computer Network Information Center of CAS. 

%\bibliographystyle{apsrev4-1}
%\bibliography{/Users/apple/MyData/Mywork/Paper/Nuclear-Phys/nuclear,/Users/apple/MyData/Mywork/Paper/Nuclear-Fission/Nuclear-Fission,/Users/apple/MyData/Mywork/Paper/Nuclear-Spectrum/Nuclear-Spectrum}
%\bibliography{nuclear,Nuclear-Fission,Nuclear-Spectrum}
%\bibliography{fission}


\end{document}